%% file: arxiv_silmarillion.tex
\documentclass[letterpaper,twocolumn,10pt]{article}
\usepackage{usenix-2020-09}

\usepackage{authblk}

\usepackage{xcolor}
\usepackage{color}

\usepackage[T1]{fontenc}
\usepackage[utf8]{inputenc}
\usepackage[linesnumbered,ruled,vlined]{algorithm2e}
\usepackage{amsthm,amsmath,amssymb}

\let\sigproof\proof\let\proof\relax
\let\sigendproof\endproof\let\endproof\relax
\let\proof\sigproof
\let\endproof\sigendproof

\usepackage{textcomp}

\usepackage{tikz}

\usepackage{graphicx}

\usepackage{booktabs}

\usepackage{tabularx}

\usepackage{enumerate}

\usepackage{url}
\usepackage{etoolbox}
\appto\UrlBreaks{\do\-}

\usepackage{hyperref}
\hypersetup{
    colorlinks=true,
    citecolor=blue,
}

\usepackage{cleveref}
\Crefname{figure}{Figure}{Figures}

\usepackage{xspace}

\usepackage{textcomp}

\usepackage[normalem]{ulem}

\def\TR{0}

\def\submit{1}

\newcommand{\sys}{Silmarillion}
\newcommand{\pancast}{PanCast}
\newcommand\covid{COVID-19\xspace}
\newcommand{\ct}{CT}
\newcommand{\spects}{SPECTS}

\newcommand{\enctr}{\textit{enctr}}

\newcommand{\skey}{\textit{sk}}
\newcommand{\pkey}{\textit{pk}}
\newcommand{\eph}{\textit{eph}}
\newcommand{\epochid}{\textit{i}}
\newcommand{\bid}{\textit{b}}
\newcommand{\did}{\textit{d}}
\newcommand{\clkinit}{\textit{C}}
\newcommand{\clk}{\textit{t}}
\newcommand{\clkstart}{\textit{start}}
\newcommand{\clkintvl}{\textit{int}}
\newcommand{\rssi}{\textit{rssi}}
\newcommand{\hash}{\textit{hash}}
\newcommand{\ephbi}{\eph_{\bid,\epochid}}
\newcommand{\attrbid}{desc_{\bid}}
\newcommand{\attrlistbid}{\{a_{\bid,1}, a_{\bid,2}, ..., a_{\bid,n}\}}

\newcommand{\etuple}{$\{\ephbi, \bid, \calc{\clk^{\clkstart}_{\bid}},
\calc{\clk^{\clkstart}_{\did}, \clk^{\clkintvl}_{\did}}, \calc{\rssi}\}$}

\newcommand{\beaconDB}{\tt BeaconDB}
\newcommand{\userDB}{\tt UserDB}
\newcommand{\riskDB}{\tt RiskDB}
\newcommand{\pirDB}{\tt D\textsuperscript{PIR}}
\newcommand{\pirDBi}[1]{\tt D\textsuperscript{PIR}[#1]}

\newcommand{\ephID}{ephemeral id}
\newcommand{\ephIDs}{eph\-emeral ids}
\newcommand{\ephmsg}{E\textsubscript{\bid,\epochid}}
\newcommand{\cf}{cuckoo filter}

\newcommand{\CF}{CF}
\newcommand{\Ltile}{L-tile}
\newcommand{\Mtile}{M-tile}
\newcommand{\Htile}{H-tile}

\newcommand{\eg}{e.g.,~}
\newcommand{\ie}{{i.e.},~}

\newcommand{\vs}{{vs.}~}

\if \submit 0
\newcommand{\todo}[1]{\textcolor{red} {#1}}
\newcommand{\update}[1]{\textcolor{black}{#1}}
\newcommand{\calc}[1]{\textcolor{magenta}{#1}}

\newcommand{\citeme}[1]{\todo{\cite{#1}}}
\newcommand{\am}[1]{\textcolor{cyan}{AM: #1}}
\newcommand{\dg}[1]{\textcolor{red}{DG: #1}}

\newcommand{\pier}[1]{\textcolor{orange}{PI: #1}}

\else
\newcommand{\todo}[1]{{#1}}
\newcommand{\update}[1]{{#1}}
\newcommand{\calc}[1]{{#1}}

\newcommand{\citeme}[1]{\cite{#1}}
\newcommand{\am}[1]{#1}
\newcommand{\dg}[1]{{#1}}

\newcommand{\pier}[1]{{#1}}

\fi

\theoremstyle{definition}

\begin{document}

\title{\fontsize{13.95}{12}\selectfont {Reconciling Security and Utility in Next-Generation
Epidemic Risk Mitigation Systems}}
\author{
  \parbox{0.95\linewidth}{\centering
  Pierfrancesco Ingo$^1$, Nichole Boufford$^2$, Ming Cheng Jiang$^2$,
Rowan Lindsey$^2$,
Matthew Lentz$^3$, Gilles Barthe$^4$, Manuel Gomez-Rodriguez$^1$,
 Bernhard Sch\"{o}lkopf$^5$,
Deepak Garg$^1$, Peter Druschel$^1$, Aastha Mehta$^2$
\\
$^1$ Max Planck Institute for Software Systems (MPI-SWS), $^2$ University of British Columbia (UBC),
$^3$~Duke University, $^4$ Max Planck Institute for Security and Privacy
(MPI-SP),\\
$^5$ Max Planck Institute for Intelligent Systems (MPI-IS)
}
}
\date{}

\maketitle

\begin{abstract}
\input{abstract}
\end{abstract}

\input{intro}
\input{overview}
\input{encounter}

\input{upload}
\input{risk-passive}

\input{implementation}
\input{eval2}
\input{discussion}
\input{related}
\input{conclusion}

%

{
\bibliographystyle{plain}
\bibliography{arxiv_silmarillion}}

\if \TR 1
\appendix
\fi

\end{document}

%% file: abstract.tex
Epidemics like the recent {\covid} require proactive contact tracing and
epidemiological analysis to predict and subsequently contain infection
transmissions. The proactive measures require large scale data collection, which
simultaneously raise concerns regarding users' privacy. Digital contact tracing
systems developed in response to {\covid} either collected extensive data for
effective analytics at the cost of users' privacy or collected minimal data for
the sake of user privacy but were ineffective in predicting and mitigating the
epidemic risks.
We present {\sys}---in preparation for future epidemics---a system that
reconciles user's privacy with rich data collection for higher utility.
In {\sys}, user devices record Bluetooth encounters with beacons installed in
strategic locations. The beacons further enrich the encounters with
geo-location, location type, and environment conditions at the beacon
installation site.
This enriched information enables detailed scientific analysis of disease
parameters as well as more accurate personalized exposure risk notification.
At the same time, {\sys} provides privacy to all participants and
non-participants at the same level as that guaranteed in digital and manual
contact tracing.


We describe the design of {\sys} and its communication protocols that
ensure user privacy and data security. We also evaluate a prototype of
{\sys} built using low-end IoT boards, showing that the power
consumption and user latencies are adequately low for a practical
deployment. Finally, we briefly report on a small-scale deployment
within a university building as a proof-of-concept.

%% file: intro.tex
\section{Introduction}
\label{sec:intro}

Containing infectious diseases, such as the recent {\covid} pandemic requires
two approaches: reactive and proactive.
Reactive measures include testing and isolating infected individuals to prevent
further spread of the disease. Proactive measures include contact tracing to
identify other at-risk individuals, and performing epidemiological analysis to
understand conditions for infection propagation, which can further inform policy
decisions.

In principle, the data required for epidemiological analysis can be collected
during contact tracing.
Unfortunately, traditional manual contact tracing does not scale well and does
not give good coverage as users tend to forget details of their recent
encounters and visits.
To scale manual tracing, several digital contact tracing systems
have been proposed recently \cite{dp3t-whitepaper, pepp-pt-ntk, tcn,
tracetogether, apple-google-tracing, bay2020bluetrace,
chan2020pact, crowdnotifier}, which record pairwise bluetooth encounters between
users' smartphones to capture physical encounters (also referred as
SPECTS\footnote{Smartphone-based Pairwise Encounter-based Contact Tracing
Systems}).
Several countries adopted centralized contact tracing systems that supported
extensive data collection for epidemiology \cite{singapore,
china-health-code-apps}. While these systems
were effective for containing {\covid}, they raised important
concerns about surveillance and users' privacy.
Other countries decided to take a more conservative approach in the interest of
users' privacy and adopted system designs that collected minimal data essential
only for contact tracing but not epidemiology \cite{covidsafe-australia,
nzcovidtracer-newzealand, swisscovidefficacy, germany-coronawarnapp,
covidalert-canada}.
\update{However, the importance of proactive epidemiological analysis can be
understood from the fact that availability of such data early on could have
helped in understanding the role of aerosols in spreading {\covid} and
enforcing social distancing and isolation much earlier
\cite{who-news,nature2022-who-delay}.}

We seek to build a secure, robust, and scalable system that expands the
utility of {\spects} by
collecting additional data relevant to future epidemics, while preserving the
privacy properties of {\spects} and manual contact tracing systems. We refer
to this as an epidemic risk mitigation system.
%
%
%
We address the following design goals in building such a system.

{\bf G1. Rich data collection:}
According to medical literature, epidemiology requires analyzing environmental
conditions, demographics, and mobility patterns that promote disease transmission
\cite{epidemiology-textbook}.
Thus, an epidemic risk mitigation system must
collect circumstantial information associated with the user encounters, such as
the location, location type, and time of encounter, as well as the environmental
conditions under which the encounters occur (\eg temperature, humidity, ambient
noise levels, etc.).
It must also support capturing non-contemporaneous encounters to determine if
a disease could transmit through indirect exposure.
Finally, the system must collect attributes of individuals (\eg age, gender,
occupation, etc.) to support
identification of vulnerable demographics.
{\bf G2. Security and privacy:}
Since the system collects sensitive user information, it
must ensure security in collection, processing, and dissemination
of the data, and balance utility and user privacy in the analytics.
{\bf G3. Timeliness:}
The system must be able to collect accurate data and disseminate risk
information in a timely manner even under a
partial deployment, low user adoption, and despite malicious or misbehaving
participants.
{\bf G4. Inclusivity:}
The system must be accessible to all demographic sections within a region of
deployment.

The effectiveness of {\spects} was also limited by non-technical factors, such
as low adoption rates. We do not address these factors in our work.

\subsection{Our solution: {\sys}}

We present {\sys}, a P2I system that relies on collection of
location/environment-tagged encounters with BLE beacons installed in strategic
locations to facilitate both contact tracing and epidemic analytics. At the same
time, {\sys} takes comprehensive measures to avoid indiscriminate collection and
dissemination of users’ encounter data, thus minimizing data leaks and misuse.

The deploying authority predetermines the analytics they wish to perform and
accordingly the set of location and environmental attributes they wish to
collect in beacon encounters. Beacons are then installed in strategic places
that may be epidemiologically relevant, such as places where people tend to
congregate (e.g., classrooms, markets, and theaters). Each beacon broadcasts (on
short-range BLE radio) identifiers called \emph{\ephIDs} that are unique to the
beacon, the current time (time is roughly quantized), and the beacon’s location
and environmental attributes (\S\ref{sec:encounter}).
Personal devices of nearby users record
these {\ephIDs} -- in particular, two users in the vicinity of the same beacon
at similar times will record the same {\ephIDs}.

When a user tests sick, the {\ephIDs} on their device from their
period of contagion are collected at a backend. Users retain full
control over what is sent to the backend (they may remove ephemeral
ids corresponding to locations they consider sensitive), all uploads
are anonymous, and a single user's ephermeral ids are divided into
small chunks that are routed separately through a
mixnet~\cite{lazar15vuvuzela} to hide the user's trajectory from the
backend (\S\ref{sec:data-upload}).



The backend periodically aggregates the {\ephIDs} uploaded by sick
individuals. It can utilize the data for epidemiological analysis,
e.g., building mobility models, detecting superspreading events,
predicting infection hotspots, determining environmental conditions
that accelerate infection transmission, etc.
\update{(The design of the analytics backend and the analytics
workloads that can be supported on the data
collected by {\sys} are beyond the scope of this work and have been covered in
other work \cite{de2022covault, roth2021mycelium}.)}

The backend also disseminates the {\ephIDs} back to everyone for {\em
decentralized} risk notification (\S\ref{sec:risk-dissemination}). Other users
match these disseminated {\ephIDs} to those stored on their own devices, and
assess their infection risk locally.

To ensure privacy of individual patients (who shared their encounter data)
during risk notification, the backend adds differentially-private noise in the
risk information disseminated to other users, which protects against a strong
adversary with auxiliary information about all other users in the system.

{\sys} also provides privacy for users when they download risk
information. Users query the backend for risk information relevant to
them using an information-theoretic private information retrieval
(PIR) protocol, without revealing their own data to the backend or an
eavesdropper.

In summary, for the desired analytics support, {\sys}'s collection of location
and environment information does not raise new privacy concerns for users. The
{\sys} backend can support rich analytics without learning sensitive information
about individual participants in the system. Privacy of sick individuals is
preserved from both the backend as well as other users and eavesdroppers.
Similarly, privacy of other users is preserved.

We build (\S\ref{sec:prototype}) and evaluate (\S\ref{sec:eval}) a prototype of
{\sys} with
battery-powered BLE beacons, and user devices ranging from smartphones to
low-end IoT devices. We demonstrate that {\sys} can be deployed with low
bandwidth, latency, and energy costs in data collection and dissemination.

To the best of our knowledge, {\sys} is the first epidemic risk
mitigation system based on P2I encounters that has actually been
implemented and evaluated.  {\sys} can be deployed incrementally by
placing beacons in locations of primary interest first.  We envision
{\sys} to be deployed as a complement to recent contact tracing
systems \cite{dp3t-whitepaper} (\S\ref{sec:discussion}). While the
latter are better suited for private or infrequently visited spaces,
{\sys} can better support crowded spaces, \update{where non-contemporaneous
transmissions may be prevalent and the existing systems would fail to
capture such events.}

%% file: overview.tex
\section{Overview}
\label{sec:arch}
\begin{figure}[t]
\centering
\includegraphics[width=\columnwidth]{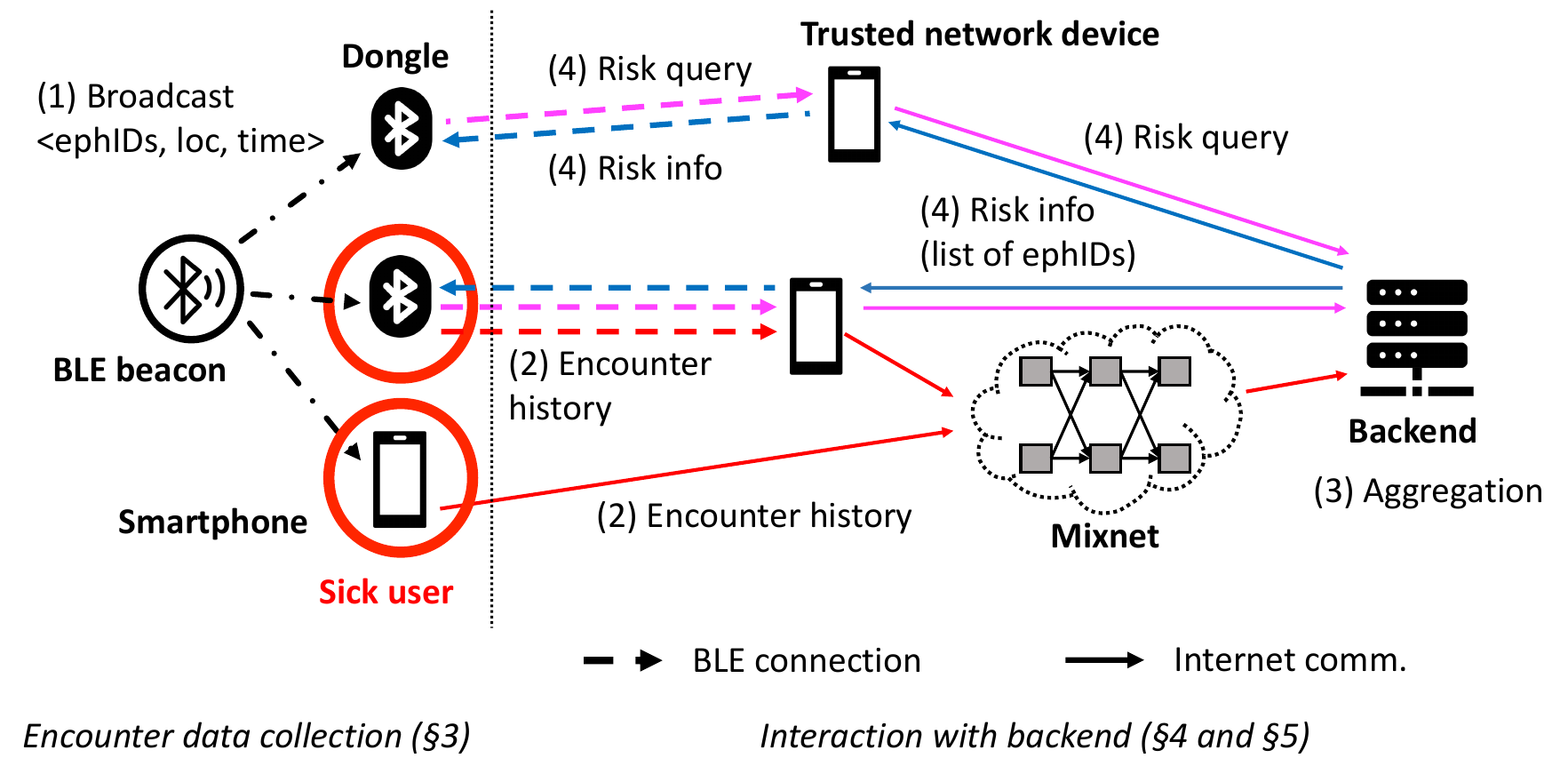}
\caption{{\sys}'s architecture and workflow.
}
\label{fig:pancast}
\end{figure}

\Cref{fig:pancast} shows an overview of {\sys}'s architecture and workflow.
{\sys}'s main components include BLE beacons, personal devices, such as
smartphones and dongles, and a backend
platform
that relays risk notifications
and aggregates data to support epidemiological analysis.

(1) The beacons are placed in strategic locations (\eg shops,
restaurants) and
continuously broadcast crypto\-graphically-gen\-erated random strings called {\em
ephemeral ids} during typical operating hours of the location. Users'
devices listen to these beacons passively (\ie without
transmitting anything) and store the beacons' {\ephIDs}.
(2) When an individual tests positive for the infectious disease, they may be
legally required to or may choose to disclose (a selected subset of) the list of
{\ephIDs} stored in their \update{personal device} to the backend. The
individual must explicitly authorize the transmission of data to the backend
from their personal device.
The user device chunks the encounter data into subsets of {\ephIDs},
packages each subset into a separate message and uploads the messages to the
backend via a mixnet.
(3) The backend
periodically (\eg daily) assimilates the information about which locations were
contaminated at which times (which depends on users' visits and location
features) into a risk database.
(4) Finally, user devices query the backend for risk information of specific
regions, compare the {\ephIDs} in their storage against those in the risk
information disseminated, and notify their owners in case of non-zero matches.

We now provide an overview of {\sys}'s components (\S\ref{subsec:components})
and threat model (\S\ref{threatmodel}).

\subsection{Components}
\label{subsec:components}
\textbf{Beacons.}
Beacons are commodity, battery-operated, BLE-capable devices that
may be installed in restaurants, squares, train stations, airports or even
mobile locations, \eg a city bus or a train. Beacons may be installed either by
health authorities or by organizations that have received an approval
from local health authorities. \update{Each day, beacons remain active during
during the typical operating hours of the location where they are installed, \ie
when the locations are visited frequently by many people.}

All beacons have a coarse-grained timer and a small flash storage.
The beacons are registered~with the backend using an id, a secret key, and
optionally a set of attributes, such as a location identifier
comprising their stationary coordinate or a route id, a region (\eg France), or
other epidemiologically-relevant descriptors about their location (\eg
humidity, temperature). The descriptors are configured statically
and can be used by the backend for
intelligent risk estimation and/or epidemiological analysis, {\em thus
addressing goal G1}.
Furthermore, beacons remain active each day during the typical operating hours
of the location where they are installed, \ie when the locations are visited
frequently by a lot of crowd. Thus, no individual user can be uniquely
identified based on their encounter with a small subset of beacons, which {\em
preserves privacy of patients sharing their data with the system and addresses
goal G2.}

\textbf{User devices.}
{\sys} enables users to participate in the system with devices ranging from
smartphones to simple dongles that can be attached to a keyring, or
worn on the wrist or around the neck.
\update{The dongles are particularly useful for physically-, technologically-,
or economoically-challenged individuals who cannot use smartphones.}

To participate in {\sys}, a user device must~minimally include a
coarse-grained timer, a counter, a small amount of flash storage, a UI to
indicate risk status and battery condition (e.g., LED), and a button to control
the LED notification.
IoT boards already offer such capabilities today \cite{silabs-dongle,
nrf52832dk, silabs-beacon} and can be used as dongles. (Smartphones naturally
have much higher capabilities.)
Dongle users require an additional trusted networked device {\em only} to upload
or download data from {\sys}'s backend. This device could be a smartphone of the
user or a care provider.

\update{By supporting diverse devices, {\sys} provides an accessible and
inclusive solution for different demographic sections of the society, {\em thus
addressing goal G4}.}
In the rest of this paper, we describe {\sys}'s design and protocols
mainly considering personal devices in the form of smartphones with a {\ct} app,
unless stated otherwise.

Similar to beacons, user devices are registered and authenticated with
the backend and receive a public-private key pair from the backend.
In addition, the users configure a password in their device, which they
use to authenticate themselves to the device. The password and
the counter are also used to control upload of data from the device.


\textbf{\update{Testing authority.}}
Users get tested at a test center which is running by a trusted authority. If a
user's test result is positive, the the test center issues a certificate for the
result signed with the center's key. Furthermore, it issues several one-time
signing keys to the user, with which the user signs their encounter entries
prior to uploading to the backend.
The signed uploads enable the backend to collect encounter data only from
diagnosed individuals and only data corresponding to their period of contagion,
thus ensuring use of accurate data for analytics and risk dissemination.
{\em Thus, it ensures security in data collection, addressing goal G2.}

\textbf{\update{Mixnet.}}
To enable users to upload their encounter entries to the backend without
revealing their identity, {\sys} relies on a mixnet similar to Vuvuzela
\cite{lazar15vuvuzela}, which mixes uploads from different users and hides the
origin of uploaded data (see \S\ref{sec:data-upload}).
{\em The mixnet ensures users' privacy during data collection, thus addressing
goal G2.}

\textbf{Backend.}
The backend may be managed by a health authority, \update{the
organization deploying beacons,} or an independent
entity. The backend maintains several databases.
(i) {\beaconDB} contains each registered beacon's
location/trajectory and the secret key used by the beacon to generate its
unique sequence of ephemeral ids.
(ii) {\userDB} contains registered users, their devices, and the
\update{public keys of the devices}.
(iii) {\riskDB} contains the encounter \update{entries} uploaded by diagnosed
individuals and is used for risk dissemination and analytics.

To facilitate risk dissemination,
the backend consists of two non-colluding servers in an IT-PIR
setup. The servers derive a PIR database out of {\riskDB}, which they use to
serve user queries for risk information of specific regions in a
privacy-preserving manner (see \S\ref{sec:risk-dissemination}).
Additionally, the backend provides differential privacy in the number of entries
uploaded by an individual patient, thus preserving patients' privacy during risk
dissemination.
{\em The risk dissemination protocol addresses goals G2 and G3}.

\subsection{Threat model}
\label{threatmodel}

{\sys} seeks to protect the privacy of users at a level~comparable to
manual tracing \update{and \spects}. The privacy of users can be
violated when they transmit information, which happens at two points
in {\sys}: (1) When sick users upload their collected ephemeral ids to
the backend, and (2) When healthy users query the backend for risk
information in regions of interest to them. {\sys} seeks to protect
the privacy of users at both these points.


Threats in {\sys} come from compromised network nodes, compromised
users and beacons, and compromise of the backend. We assume a standard
network adversary that can compromise a subset of the network nodes
(routers, switches, servers), and monitor all traffic on the
compromised nodes, but it cannot compromise a significant fraction of
network nodes. This is particularly important for our use of a mixnet,
where we assume that at least one mixnet node is uncompromised.

For users and beacons, we follow a \emph{mostly honest} model, where
users and beacons are generally honest but a small fraction may be
controlled to act arbitrarily by the adversary. The backend is assumed
to be honest-but-curious; if compromised, it follows the prescribed
protocols but it may try to use information it sees and information
that compromised users, beacons and network nodes see to break user
privacy.

\emph{Side channels.}
Side channels (\eg EM, power), which could be exploited to steal
devices' crypto keys, are out of scope.
In practice, devices could implement constant-time
crypto~\cite{bear-ssl-constant-time} to mitigate these attacks.

%% file: encounter.tex
\section{Encounter data collection}
\label{sec:encounter}


%
In this section, we describe the device configurations, the interactions of
user devices with beacons, and the collection of encounter data on user devices.
In \S\ref{sec:data-upload}, we discuss how user devices upload the
data to a backend to facilitate epidemiological analysis and subsequent risk
dissemination.
In \S\ref{sec:risk-dissemination}, we discuss how user devices interact with the
backend to receive the risk information for contact tracing.

\subsection{Initial configuration}
\label{subsec:pancast-config}

Prior to its installation, a beacon is
configured with a unique id $\bid$, an initial clock $\clkinit_{\bid}$
synced to real time, a secret
key $\skey_{\bid}$ that is known only to the beacon and the
backend, and an optional
descriptor $\calc{\attrbid = \attrlistbid}$ that includes attributes, such
as a location id, environmental conditions, indoor/outdoor, average temperature,
ventilation or other important features of the place where the beacon will be
installed.


Each \update{user device} is configured with a unique id $\did$, the backend's
public key, a initial clock $\clkinit_{\did}$ synced to real time, a
public-private key pair $(\pkey_{did}, \skey_{\did})$,
\update{a monotonic counter $ctr$ from the backend, and a password from~the
user.}
\update{In smartphones, the initial clock value may be the device's own wall-clock time. In dongles, the initial clock is set to the backend's wall-clock time at
the time of device registration.}
%
The \update{device's} \update{initial counter value} is known only to the
device and the backend and is used to ensure freshness of uploads from the
device to the backend. The secret key $\skey_{\did}$ is stored only in the
device and never leaves the device.
The~\update{password is}
used to mutually authenticate the owner and the device whenever the owner
interacts with the device (\eg to initiate upload to the backend as in
\S\ref{sec:data-upload}).
\update{For dongle users, the password is configured in both the dongle and the
trusted networked device.}




The beacon configurations are registered with the backend, which
stores this data in the {\beaconDB} database in the form:
\mbox{\{\textit{device~id}, \textit{device key}, \textit{initial clock},
\textit{clock offset}, \calc{\textit{descriptor}}\}}, where \textit{device key}
is the secret key for a beacon.

Similar to beacons, user device configurations are also registered with the
backend and stored in the {\userDB} database in the form:
\mbox{\{\textit{device~id}, \textit{device key}, \textit{initial clock},
\textit{clock offset}},
where \textit{device key} is
the public key for a user device.

The clock offset in the backend is initialized to 0 during registration of a
device and later used to track any divergence between the
real time and the local timer of the device.

\subsection{Capturing beacon encounters}
\label{subsec:capture-beacon-encounters}
Each \update{user device} and beacon has a coarse-grained timer of 1-minute
resolution
($\clk_\did$ and $\clk_\bid$ respectively), which is set to the initial clock
value provided by the backend, and subsequently increments every minute. A
device stores its timer value to local storage at intervals of fixed
length $L$, called epochs. A variable tracks the epoch id or the
number of epochs elapsed since the device's start ($\epochid_\did$ and
$\epochid_\bid$ for a \update{user device} and beacon, respectively).
In our prototype, we use epoch length $L=15~\text{minutes}$, similar to
 recent {\ct} systems.

A beacon generates a new ephemeral id every epoch. In the $\epochid^{th}$ epoch
$\epochid_{\bid}$, the beacon $\bid$ generates an id
$\eph_{\bid,\epochid} = \hash(\skey_{\bid}, \epochid_{\bid}, \calc{\attrbid})$.
Here $\epochid_\bid = \lfloor (\clk_\bid - \clkinit_\bid)/L \rfloor$ and $\hash$
is a one-way hash function.
\update{The beacon broadcasts ${\ephmsg} = \{\ephbi, \bid, \epochid_{\bid}\}$ on
legacy BLE advertisement channel and its descriptor $\calc{\attrbid}$ on a
separate periodic advertising channel.}
%
Beacons broadcast each {\ephmsg} several times within an epoch.
When a \update{user device} is in the bluetooth range of a beacon, it captures
the beacon broadcasts. \update{If the device encounters a new beacon id, it
briefly listens to the periodic advertisement channel of the beacon to
additionally capture the beacon's descriptor once.}
A \update{user device} persists a single entry for each unique {\ephID}
along with the
first beacon and \update{device} timestamps at which the id was received
($\clk^\clkstart_\bid$ and $\clk^\clkstart_\did$), the duration for which the id
was observed ($\clk^\clkintvl_\did$), and the
average of the RSSI values observed ($rssi$).
Thus, a log entry $\enctr$ in user device database $\did$ would be: \etuple.
\update{The device stores one instance of each unique $\attrbid$ captured, which
it may use to provide the device owner descriptive information about the owner's
trajectory.}

%
\textbf{Datastructure configurations.}
The byte size of each field in an ${\enctr}$ is as follows. The {\ephID}
$\ephbi$ is \calc{23} bytes, the device id
$\bid$ is 4 bytes, the timestamps $\clk^\clkstart_\bid$, $\clk^\clkstart_\did$
and the interval $\clk^\clkintvl_\did$ are 4 bytes each, and $\rssi$ is 1
byte.
The ${\ephbi}$ is generated by computing a SHA-256 hash of the
inputs and taking the least significant \calc{23} bytes of the result.
Each beacon broadcast {\ephmsg} is \calc{31} bytes and fits in a single legacy
BLE advertisement; the descriptor $\attrbid$ can have variable length.
Each encounter stored in a \update{user device} is \calc{40 bytes}.

Assuming that users encounter~on average no more than one unique {\ephID}
every \calc{10 min} in a day, \update{user devices} need to store data for
\calc{2016} encounters in a \calc{14-day} window (the infectious period for the
\covid disease as determined by health experts), which requires
\calc{$\sim$\mbox{79 KB}} of persistent storage.
\update{Assuming that each encounter also corresponds to a unique beacon and an
average beacon descriptor length of 64 bytes, the device requires an additional
126 KB of persistent storage for the descriptors.}
In reality, a \update{device}~is~unlikely to encounter a unique {\ephID}
every \calc{10 min}, much less a unique beacon, continuously for 14 days.
Hence this estimate~is very conservative.
\update{Overall, the storage requirement is satisfied by both smartphones and
many IoT devices.}

%% file: upload.tex
\section{Encounter data upload}
\label{sec:data-upload}

When users feel sick or are notified of potential exposure
(\S\ref{sec:risk-dissemination}), they may visit a test center or a clinic for
testing.
%
Patients identify themselves at the time a test is taken using the
normal procedures in place for this purpose. Normally, their contact details are
recorded along with the id of \update{their device and} the test kit used for
them. Once the test results are available, the user is informed using their
contact details, such as their email id or phone number.
If the result is positive, the user may wish to or be required by law to upload
their data to assist in dissemination of risk information and epidemiological
analytics.

We start by discussing the competing challenges involved in
designing a privacy-preserving upload mechanism.

\subsection{Requirements}
The upload mechanism needs to address four requirements.
First, because encounters contain contextual information (\eg beacon
location), uploading encounters may reveal a user's entire
trajectory, which would be a violation of their privacy.
To ensure privacy of diagnosed individuals, the
upload protocol must provide:
\begin{itemize}
    \item[{\bf U1.}] {\em Anonymity}: the backend or a network adversary cannot
    learn the identity of any user uploading encounters.

    \item[{\bf U2.}] {\em Unlinkability}: the backend or an adversary cannot
    learn if two parts of a trajectory belong to the same user or~not.

\end{itemize}
Furthermore, the protocol must be reasonably efficient in terms of
overall network traffic:
\begin{itemize}
    \item[{\bf U3.}] {\em Efficiency}: The network traffic generated
      by the protocol should be linear in the amount of data actually
      transferred from users to the backend, ideally higher by only a small
      factor.
\end{itemize}
Finally, the protocol must be robust against malicious users who may attempt to
generate false alarms and panic among users, for instance, by uploading fake
entries to the backend or uploading legitimate entries without having been
diagnosed positive. Specifically, the protocol must support:
\begin{itemize}
    \item[{\bf U4.}] {\em Upload authentication}: the backend must verify
    that the uploads came from a registered user who tested positive.
\end{itemize}

We discuss {\sys}'s mixnet-based upload protocol in
\S\ref{subsec:upload-protocol}, which addresses the requirements U1-U3. We
discuss authentication (U4) in \S\ref{subsec:upload-auth} and initiation of
encounter uploads from user devices in \S\ref{subsec:device-upload}.
%


\subsection{Upload protocol}
\label{subsec:upload-protocol}

\textbf{Message format.}
Since uploading a user's complete encounter history can compromise the user's
privacy, the user device chunks the history into small subsets of
$t$ encounter entries and uploads them in separate messages.
The privacy guarantees rely on a key assumption that a user cannot be uniquely
identified by a small segment of $t$ {\ephID} records of her trajectory. This is
a reasonable assumption since we expect {\sys} beacons to be installed
strategically in crowded places during busy hours, \eg train stations, airports,
markets, etc. (see \S\ref{sec:arch}).

A user device splits the encounter data into messages as follows. First, it
shuffles the encounter entries in the device log, and then divides
the shuffled log into $24$ subsets (one for each hour of the day).
Each subset contains at least $t$ and at most $\calc{2016 / 24 = 84}$ {\ephIDs}.
(Recall from \S\ref{sec:encounter} that the max number of entries in a device
log can be 2016.)
Then, the device places each subset of entries in a separate
message, pads each message with dummy entries as required up to a fixed message
size $M$, encrypts the message with the backend's public key \update{and signs
the message with a unique key provided by the test center (see
\S\ref{subsec:upload-auth} for signing messages).
Each message
is then uploaded to the backend through a mixnet, as explained below.}

\textbf{Mixnet rounds.}
{\sys}'s upload protocol relies on a mixnet, such as \cite{lazar18karaoke},
\cite{corrigan10dissent}, \cite{lazar15vuvuzela}.
We assume the mixnet consists of a chain of $r$ servers.
\update{We make the standard assumption about the mixnet service that at least
one server in the mixnet is honest.}

The upload protocol runs in \update{synchronous} rounds.
Specifically, we divide each hour into $n_r$ rounds, each of them $60
/ n_r$ minutes long.  \update{Every hour,} each user device sends
messages to the mixnet in only one of those rounds. Each device is
assigned a round randomly. A device uploads (a subset of) the
encounter data in the message if available, and dummy data otherwise.
%

We now explain how \textbf{U1--U3} are attained.

\textbf{U1}: Given $n$ {\sys} users, and an average participation rate
of $R$ in any given round, the probability that any given round has
$k$ participants is $\binom{n/n_r}{k} R^k (1-R)^{(n/n_r) - k}$. Even
for a small city with $n = 100,000$, and $n_r = 10$, $R = 10\%$, the
probability that there are at least 1,000 participants in a given
round is more than $50\%$, which implies a high degree of anonymity.

\textbf{U2}: Users anonymize small subsets of their trajectories and
beacons operate only in densely visited places, making it difficult
for the an adversary to link two trajectory subsets to the same user.

\textbf{U3}: The average network traffic generated every round in the
system due to the upload protocol is
\begin{equation*}
	T = \frac{n \cdot R \cdot M \cdot r}{n_r}
\end{equation*}
If a user is sick with probability $p$ on any given day, then the
actually meaningful traffic would be $A = (n/n_r) \cdot R \cdot M
\cdot p$ per round. Hence, the average traffic overhead is $T/A =
r/p$. For $r = 4$ (a 4-round mixnet) and $p = 0.02$ (2\% users sick at
any given time), this overhead is 200x. Given that the actual traffic
generated by each sick user's device is about 126KB in 14 days or 9KB
per day (see \cref{subsec:capture-beacon-encounters}), this 200x
overhead still amounts to only 1.8 MB traffic per device each day,
which can be easily tolerated even when user devices have limited
network connectivity.



\subsection{Upload authentication}
\label{subsec:upload-auth}
We now describe how a user device authenticates and initiates uploads.
%
%
One concern for uploading is that users may upload incorrect or fake encounter
entries, \eg by uploading entries without having been diagnosed positive or by
uploading entries that are older than the period of contagion.
\update{The backend can easily discard dummy and invalid entries that could not
have been generated by any registered beacons, as well as entries with
timestamps that are too old to be relevant for risk notification or
epidemiology.}
To mitigate the risk of users uploading without being sick, we describe a
mechanism to authenticate user uploads.

An upload authentication mechanism must enable the backend to verify
that each entry has been uploaded by a user who was diagnosed positive
by an authorized test center.  A simple solution would be having users
upload their encounter data along with a certificate from the test
center, signed with the test center's key, indicating the test date,
and the ids of the patient's device and test kit. However, the user
would need to upload the certificate with each encounter entry, which
would defeat the goal of ensuring unlinkability of the user's
entries. Instead, the user must be able to attest each of their
encounter entries independently. We describe the solution next.

The authentication mechanism relies on test centers playing the role
of a trusted third-party.  When a test center generates a positive
result for a user, it releases a one-time password (OTP) to the
diagnosed user and to the backend. The OTP may be derived from a
master secret $M_T$ of the test center and a counter $C_T$
representing the number of users who tested positive at the
center. The user then sends the OTP to the backend and downloads $N$
one-time signing keys from a database of keys in the backend using an
\emph{oblivious transfer} (OT) protocol, which prevents the backend
from learning sets of keys that were downloaded
together. Subsequently, the user can sign each of their trajectory subsets with
one of the downloaded keys each. When the user uploads
the subsets, the backend can verify the signatures on the subsets by
trying the verification keys. Since the backend does not know which
keys were given to the same user, it cannot link the different uploads
of the same user to each other.


The authentication mechanism can partially prevent a sick user from
authenticating arbitrary entries and uploading them to the backend.
A sick user may authenticate entries that their device never recorded, \eg by
copying entries from a colluder's device to their own device. The backend may be
able to detect if a user uploads entries from multiple distant locations at the
same time; however, it can do so only within a subset of entries uploaded
together, but not across independent subsets. Thus, the length of the trajectory
subsets trades off unlinkability and the ability to detect malicious behavior.

A malicious user could also simply upload a consistent encounter history of a
different device. This risk cannot be entirely
eliminated, since it is difficult to verify whether a device logged
entries in the proximity of beacons or not. However, this {\em analog
    loophole} exists in all digital contact tracing systems, not just the
one we are proposing here.

\subsection{Initiating upload from a user device}
\label{subsec:device-upload}
Next, we discuss how user devices initiate upload of entries.
Depending on the~user device, the upload mechanism requires different steps as
described below.

\textbf{Smartphone upload.}
A user initiates the data upload after they receive a positive test result.
A smartphone user downloads the OTP from the test center, forwards it to the
backend and downloads the one-time signing keys from the backend, all over the
internet.

\textbf{Dongle upload.}
Dongle users need to download the signing keys with the help of a trusted
network device. The trusted device could download the OTP from the test center
on behalf of the dongle, forward it to the backend, download the one-time
signing keys from the backend and finally forward the keys to the dongle over
BLE.
To initiate upload, the dongle user then establishes a secure connection between
the dongle and the trusted device by entering their dongle's id and password on
the trusted device's UI and instructing the device to establish an authenticated
session with the dongle.
The dongle encrypts each encounter entry
with the backend's public key, signs it with one of the signing keys, and then
uploads the encrypted and signed payload to the trusted device, which then
uploads entries to the backend via the mixnet.

Note that the encounter history is not released in cleartext to the
personal device.

\subsection{Security analysis of the upload mechanism}
The upload protocol authenticates a user's encounter entries to the backend
without linking the trajectory subsets to each other or to the user. This is
achieved as
follows. If a backend can verify the signature on an uploaded trajectory subset,
it knows
the subset was signed using a one-time signing key provided by the backend. A
user could have received the signing key only upon authenticating itself to the
backend with a valid OTP that was generated by a trusted test center, which in
turn would have generated the OTP only if the user was diagnosed positive.
At the same time, the OT protocol prevents the backend from learning the
keys downloaded by the user and the user from learning the keys that they do not
wish to use.

Even when the backend has retrieved all the {\ephIDs} that were deposited
in all the mailboxes, it cannot piece together the full trajectory of any single
user. This is because, by assumption, no {\ephID} is unique to any individual
and the uploading protocol prevents linking a sequence of messages together and
to a specific individual. Moreover, the backend cannot identify the sick users
and, therefore, cannot know with certainty to which user a particular set of
{\ephIDs} belong.

\update{Nevertheless, for added protection of the encounter data collected,
the backend can be further secured using standard hardware and cryptographic
techniques~\cite{de2022covault}.}

%% file: risk-passive.tex
\section{Risk dissemination}
\label{sec:risk-dissemination}


We start with an overview of the risk information structure and the risk
notification mechanism in the \update{user devices}.
The risk~information consists of a list of {\ephIDs}. The {\ephID} of a
beacon $b$ for epoch $i$ is included in the list only if a diagnosed
individual encountered $b$ in epoch $i$.
For accurate~risk estimation, the risk information
may contain additional encounter parameters, \eg \calc{rssi, encounter duration,
beacon's descriptor}, and weights for the beacon descriptors.

If a \update{user device} has previously recorded any of the {\ephIDs} listed in
the risk information, its owner may have been exposed to a diagnosed individual.
The \update{device} computes a risk score based on the number of matched
{\ephIDs} and (optionally) other features of the matched encounters.
If the risk exceeds a certain threshold, the \update{device} notifies the user
so that they can self-isolate and get tested.
\update{In dongles, the notification can be generated by having the user press a
button on the dongle and the LED blink with a specific pattern.}

We now discuss how the risk information is disseminated from the backend
to \update{user devices}.
We assume that most users check their risk status once a day on average. We
start with the requirements that the risk dissemination protocol needs to
satisfy and then describe how {\sys} satisfies each of the requirements.

\subsection{Requirements}
%
The risk dissemination protocol needs to address four requirements, which we
discuss in this section.
{\bf D1.} the information disseminated must be correct,
{\bf D2.} the protocol must preserve the privacy of the diagnosed patients whose
information is being disseminated (\S\ref{subsec:risk-patients}),
{\bf D3.} the protocol must maintain privacy of the users seeking the risk
information (\S\ref{subsec:dissemination-protocol}), and~{\bf D4.} the relevant information must reach potentially affected users
in a timely manner and with low bandwidth, power, and computational
costs for \update{user devices} (see \S\ref{subsec:dissemination-protocol}).
%
Note that all risk information is signed by the backend to allow detection of
any tampering, which addresses D1.

\subsection{Noising the risk dissemination}
\label{subsec:risk-patients}


We present two scenarios where the \emph{number of entries} in the risk
information could potentially reveal an individual's \update{movements}
or health status to an adversary in the locality of the individual. We
then describe our solution to mitigate such leaks, addressing
requirement~D2.

{\em (i) \update{Movements} of diagnosed individuals.}
Suppose Alice learns (from the local news) that there was only one case of
infection in the past few days within some geographic region.
Separately, she learns that Bob was diagnosed and that he agreed to upload his
encounter history when he got diagnosed.
Alice can infer if Bob~was near any beacon in the region while he was
contagious based on whether she receives risk information for the region or not.
Thus, the length of the risk information (zero \vs non-zero) reveals to an
adversary information about the \update{movements} of a diagnosed individual.


{\em (ii) Health status of an individual.}
%
Suppose Alice lives in an area with few people, say $n$, and Alice
is~able to track the movements of $n-1$ of these people through outside
channels. If Alice receives risk information with more {\ephIDs}
than can be accounted for by the movements of the $n-1$ people she is
tracking, she knows that the $n$th person (whom she is not
tracking) must be sick as well.
Even though such an attack requires a significant amount of offline information
and may be difficult in practice, it does raise privacy concerns.

Note that these leaks rely solely on the \emph{number of
ephemeral ids in a risk notification} and arise without~the adversary
having even encountered an individual.
We mitigate these leaks by adding noise to the risk information to
hide the actual number of {\ephIDs}. We~add junk ids that do not
correspond to any real beacon and thus do not match the history of
any user \update{device}.
Given our threat model, no adversary can monitor the ephemeral ids
from a significant fraction of beacons and, thus, 
distinguish the junk ids from legitimate~ids.
The number of junk ids satisfy differential privacy (DP), which we describe next.

We adapt a mechanism proposed in prior work~\cite{DBLP:conf/ccs/AkkusCHFG12}.
%
Given a risk payload, we add $N$ junk ids to it, where
$N = t + \lfloor \tilde{X} \rfloor$ is always non-negative;
$t$ is a natural number, and $\tilde{X}$ is a random~value sampled from a
Laplacian distribution with mean $0$ and parameter $\lambda$ \emph{truncated} to
the interval $[-t, \infty)$. The values of
$t$ and $\lambda$ depend on the privacy required. To
get $(\epsilon, \delta)$-DP, we pick $\lambda = A/\epsilon$ and
$t = \lceil \lambda \cdot \ln \left( (e^{(A/\lambda)} - 1 + \delta)/2\delta
\right) \rceil$.
Here, $A$ is the sensitivity of the risk payload function; it equals the
maximum number of risk entries that could be contributed by a \emph{single}
diagnosed individual, which we conservatively set to \calc{2016}
(\S\ref{subsec:capture-beacon-encounters}).
For $\epsilon = 0.1$ and $\delta = 0.01$, the 99th \%ile noise required
is 115991.
\if \TR 1
We prove that our mechanism is $(\epsilon, \delta)$-DP in appendix \ref{sec:appendix-dp}.
\else
We prove that our mechanism is $(\epsilon, \delta)$-DP in an extended tech
report~\citeme{silmarillion-techreport}.
\fi

%
%

\subsection{Dissemination protocol}
\label{subsec:dissemination-protocol}
A key requirement for {\sys} is to ensure that users can receive risk
information without revealing their own encounter history.
Additionally, traveling users must be able to access global risk information
to receive reliable risk estimation.

A na\"ive way to satisfy these requirements would be to broadcast the complete
risk information to users and let the users' devices filter the data for
relevant matches.
However, this could incur high latency, power,
and computational costs for the \update{user devices}.
\update{{\sys} simultaneously addresses requirements D3 and D4 by using a IT-PIR
protocol that allows users to query the backend for risk information without
revealing their own encounter entries.
}\update{Below, we describe the PIR datastructures and the
protocol.}

\textbf{PIR datastructure.}
The backend enables users to query for risk information in fixed-sized blocks,
where blocks are derived by grouping the ${\riskDB}$ entries based on a set of
one
or more beacon attributes.
The grouping function $G$ may be,
for instance, based on a region identifier attribute associated with the beacons
(\eg zip code or country code), or based on a location type attribute (\eg
airports, theatres, etc.). The backend may support one or more
grouping functions.
Suppose a function $G$ yields $N$ possible values for its attribute set $\{g_1,
g_2, ..., g_N\}$. The backend maintains an $N$-length array ${\pirDB}$,
where entry ${\pirDBi{i}}$ corresponds to a block for $g_i$. The entry is
a data block if
there are non-zero number of ${\riskDB}$ entries with $g_i$ in their beacon
attributes, otherwise it contains a dummy block.
In other words,
\begin{equation*}
    \pirDBi{i} =
    \begin{cases}
        G(\riskDB, g_i), & \text{if}~\exists~e \in \riskDB,~g_i \in e.desc_b\\
        \text{dummy}, & \text{otherwise}
    \end{cases}
\end{equation*}

For each function $G$, the backend maintains a separate ${\pirDB}$.

To ensure that a user does not learn the actual number of {\ephID} entries
within a block (\eg in a region-based grouping) and to make the block size
uniform, the backend adds
dummy entries to each block, following the DP mechanism of
\S\ref{subsec:risk-patients}, upto the uniform block size of ${\pirDB}$.

\textbf{Dynamic and hierarchical grouping.}
A key practical challenge is that as the entries uploaded by users evolve each
day, the distribution of risk entries may vary in the groups generated by a
grouping function. Consequently, the block size for a ${\pirDB}$ may need to be
changed frequently. Furthermore, a skewed distribution of risk entries may
require very large block sizes, leading to unnecessary bandwidth overheads.
Therefore, the backend dynamically adjusts the grouping of {\riskDB} entries
based on the prevailing infection rate distribution, while maintaining a uniform
block size~$B$.

For a grouping function, the backend organizes the attribute ids hierarchically
like a B\textsuperscript{+}-tree, whose height and fanout depend on the
desired block size.
A ${\pirDB}$ block corresponds~to a B\textsuperscript{+}-tree node at a certain
level if the total
number of risk entries for all nodes below it is $\leq B$. When the number of
entries overflows $B$, the entries are split into new ${\pirDB}$ blocks that are
associated with B\textsuperscript{+}-tree nodes at the next lower level.
Partially-filled blocks are padded with dummy entries as above.

\begin{figure}[t]
\centering
\includegraphics[width=\columnwidth]{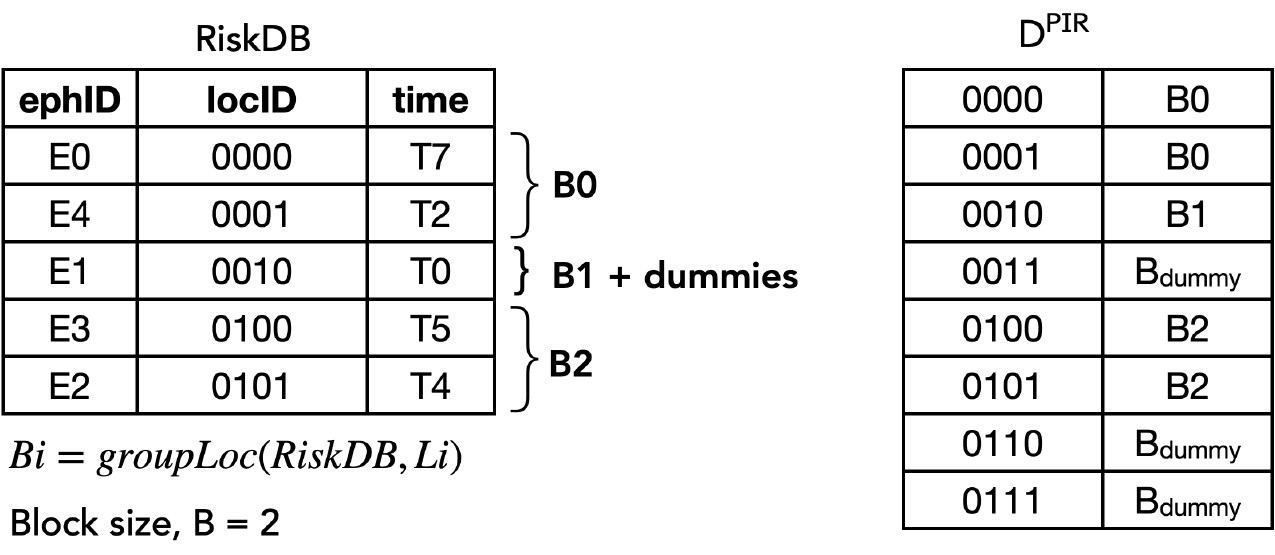}
\caption{
  PIR DB in the backend on a given day.
}
\label{fig:pir}
\end{figure}

\textbf{Block encoding.}
{Once the blocks are generated, the backend then encodes
each block into a {\cf} ({\CF})~\cite{fan2014cuckoo}.
The {\CF} encoding ensures that users~can only check for
presence of~specific ephemeral ids but not learn all the ids in
the risk payload, thus slowing down {\ephID} harvesting by
colluding users. The {\CF}s also reduce bandwidth overheads, albeit at
the cost of a small percentage of false positives. For instance, a {\CF}
with entries of size \calc{32 bits} (as opposed to the 15-byte \ephIDs) reduces
risk payload sizes by  \calc{$\sim$3.75x} while incurring 
\calc{<0.01\%} of false positives for a 14-day period.}
{\Cref{fig:pir} shows the PIR database.}


\textbf{PIR protocol.}
We use a PIR protocol based on~\cite{chor1995private}.
Our scheme relies on two servers, $S_1$ and $S_2$, each of which has a complete
copy of the ${\pirDB}$.
We make the standard assumption that at least one server is non-malicious
and non-colluding.
Suppose ${\pirDB}$ contains $N$ blocks and let the size of the largest block
in ${\pirDB}$ be $B$ bits. We
assume that all blocks are padded with dummy entries up to size~$B$.

First, the user device queries the backend for the grouping functions supported.
In {\sys}, we support the region-based grouping function which maps each
encounter entry to an enumerated region id. The user device applies the function
on its own encounter log to determine the unique regions visited and,
therefore, for which it needs the risk information.

To query for the $n$-th region (block) in ${\pirDB}$, a \update{user device}
generates two secrets shares $Q_1, Q_2$, which are random bit strings of
length $N$ with same bits except~the
$n$-th bit, which is flipped.
The \update{device} sends $Q_j$ to $S_j$; $j$~$\in$ \{1, 2\}.

Each server expands its query share from a vector of 1-bit entries to a vector
of $B$-bits entries by replicating the bit at each index in
the original query $B$ times.
Next, each server $S_j$ generates a response
$A_j = \bigoplus_{k = 0}^{N} (Q_j[k] \cdot \pirDBi{k})$,
encrypts it with the client device's encryption key, and returns it to the
client.
The \update{user device} decrypts each response from the two servers and XOR's
them to retrieve ${\pirDBi{n}}$.

The \update{user device}'s complexity for generating a query-pair and uploading
the queries, and each backend's computation complexity are $O(N)$ each. The cost
for receiving the response shares and retrieving a block is $O(B)$ each.

For querying, the \update{user device} \update{sets up an end-to-end secure
session with each backend PIR server and then issues PIR queries for each unique
block covering the logged encounters.}
\update{A smartphone user can directly connect with the servers over Internet. A
dongle user offloads risk querying to its trusted networked
device}, as is done for encounter history upload.

Furthermore, the \update{user device} generates fake queries using DP (similar
to \S\ref{subsec:risk-patients}) to hide the actual
number of queries issued in each round of risk querying (about once a day).

\subsection{Security analysis of risk dissemination}
Users learn nothing about other users except through the risk
notifications, and the only information they can learn is that which
is uploaded by diagnosed users.  Thus, it is impossible~to learn
anything about healthy users who never share any information with the
system.

For diagnosed individuals, the DP mechanism in the risk dissemination protects
the number of encounter entries uploaded by them, while the {\cf} encoding of
risk information prevents others from easily learning the actual entries.

{\sys} also ensures strong privacy of users receiving risk information.
As long as one of the PIR servers is non-malicious,
user's privacy is protected from the servers, since each server receives only a
share of the query and iterates over the entire ${\pirDB}$ regardless of the
query, and only the user can recover the DB block from the response shares of the
servers.
User generates a number of queries following DP, and all queries and results are
encrypted end-to-end between the clients and servers, thus preventing leaks to
eavesdroppers.
The only remaining leak is through the timing and number of risk notification
rounds.

Thus, the backend learns no information about queriers, particularly healthy
users, except the times when they query for risk data.
%

\textbf{Limiting case.}  A user can still learn the specific {\ephIDs} it
collected from a beacon they visited at the same time as a sick individual.
Combining with auxiliary information (\eg from local
news), she might be able to isolate the location trajectory of an
individual and ultimately identify an individual in the worst
case.
Note that {\em such leaks are inherent to all digital tracing systems}.
Nonetheless, DP
(\S\ref{subsec:risk-patients}) combined with our assumption that an
adversary cannot record {\ephIDs} widely ensures that users cannot
learn the complete history of an individual without stalking them
physically.

%% file: implementation.tex
\section{{\sys} prototype}
\label{sec:prototype}

As a proof of concept, we implemented {\sys} using low-cost BLE beacons,
BLE-only dongles, and smartphones with both BLE and network capabilities.
We implemented BLE beacons on Nordic nRF52832 development kits~\cite{nrf52832dk}
and BLE dongles on SiLabs Thunderboard Kit SLTB010A~\cite{silabs-dongle}. All
the Bluetooth devices support BLE~5.0.
The BLE-only devices are powered through 3V/220mAh CR2032 coin cells.
Only 64KB of flash storage is usable in dongles, with the remaining storage
being used by the platform software. While the storage was sufficient
in our evaluation (\S\ref{sec:eval}), we
recommend using devices with at least 128KB of log storage.

\textbf{BLE beacons.}
The BLE beacons are configured with a location id attribute, which is an
8-bit integer indicating geo-coordinates of the beacon within a 1
km\textsuperscript{2} region.
The beacons
transmit {\ephIDs}
as legacy advertisements on BLE's advertising
channels. 

\textbf{Dongle.}
The dongle's key datastructure is a circular log on flash, which is used to
store records of beacon encounters.
%
The dongle implements four event handlers.

An {\em encounter handler} scans on the legacy advertisement channel at 1s
intervals with a duty cycle of 10\%.
When the dongle receives a packet, the handler decodes the encounter
payload. It adds a new encounter to an in-memory list of "active" encounters, or
updates an existing encounter's interval since the first instance of the same
encounter was observed.

A {\em clock handler}, triggered once per minute, increments the dongle's clock
and performs different actions on encounter entries in memory and on
store depending on their state. It deletes stored encounters older than 14 days.
For the "active" encounters 
older than one epoch (and thus ready to persist in the encounter log),
it computes their cuckoo filter lookup indexes 
and appends the encounter and the indexes to the log.

A {\em query handler}, triggered on a button press, initiates the querying
protocol for risk information. The dongle aggregates regions to query
from the location ids of the recorded encounter entries. For each query, the
dongle generates two secret shares using a hardware TRNG and encrypts each
share using a separate key generated with hardware AES-CBC256.
The keys can be derived from \update{the dongle key and counter}~pre-shared with each backend PIR
server. The dongle then sends both queries to a network beacon over a BLE
connection, and the beacon then forwards the queries to the respective server.
Each server encrypts its PIR response with its AES key and sends the response
to a network device, which then transmits the response to the listening
dongle. The dongle decrypts and~XORs the response shares to retrieve the final
response, decodes the cuckoo filter in the response,
looks up each dongle log entry in the filter, and sets a bit for each matched
entry.

An {\em LED handler} periodically toggles a device LED. Normally, the LED blinks
5 times at 1s intervals every 2~min to indicate that the device is
alive.
After downloading risk entries, the user can press a button to check for an
exposure (new matches in {\CF}), which is indicated by the LED blinking
continuously at 0.25s intervals for 2 min before resetting to the
normal rate.

Due to limited RAM, dongles compute risk scores in a streaming
manner. The query handler downloads a chunk of risk payload,
performs the necessary lookups, then discards the chunk before downloading
another chunk. The chunk ids track pending chunks.

Although straightforward, we did not implement the upload pipeline from a dongle
to a network device, since the costs of this pipeline would be much
smaller than the costs for~the network device to participate in the mixnet
protocol for uploading to the backend.

\textbf{Smartphone.}
We also implemented the \update{user device} functionalities as an Android~11
app. The app captures beacon encounters similar to the dongle. Additionally, it
uploads the device's encounter history for the last 14 days to the backend over
HTTPS and downloads the
complete risk data of the last 14 days from the backend over HTTPS.

\textbf{Backend.}
The backend server runs on \update{two} Dell PowerEdge R730 Servers,
each with 16 Intel
Xeon E5-2667, 3.2GHz cores, 512 GB RAM, and 1TB SSD.
It maintains ${\pirDB}$ as an in-memory array,
uses AVX256 for PIR, and computes
new {\CF}s daily from the uploaded {\ephIDs} 
of the last 14 days.
\update{For our experiment, we use a PIR block size of 5 MB and a PIR database
$\pirDB$ of \textasciitilde430K blocks (regions).}
%
To enable incremental risk score updates in \update{user devices}, the backend
splits the data into multiple {\cf}s, which are transmitted as chunks with
distinct ids.
\update{Each chunk includes a filter of 128 indices with 4 buckets per index.}
The backend only transmits {\ephIDs} in the risk information; extensions for
intelligent risk estimation are left for future work.



%% file: eval2.tex
\section{Evaluation}
\label{sec:eval}

\begin{table}[t]
\begin{tabular}{p{0.24\columnwidth}p{0.15\columnwidth}p{0.14\columnwidth}p{0.16\columnwidth}p{0.05\columnwidth}}
\toprule
{\bf Operation} & {\bf Frequency} & {\bf Compute} & {\bf Bandwidth} & {\bf UI}
\\
\midrule
ephid collect & high & low & low & no
\\
risk calc & med & high & high & low
\\
history upload & rare & med & med & med
\\
\bottomrule
\end{tabular}
\caption{Dongle operations. (UI = User involvement)}
\label{tab:dongle-ops}
\end{table}

In this paper, we evaluate the practicality and usability of {\sys}'s design and
implementation for the use case of contact tracing and risk notification.
An evaluation of the effectiveness of beacon-based tracing has been shown
in prior work~\cite{pancast-simulations}.


BLE beacons perform a single task of generating {\ephIDs} periodically; thus,
they require low maintenance and the only practical concern is their battery
life.
For users' devices, the practicality and usability is determined by the
frequency of interactions required with the devices,
the timeliness of risk dissemination, the bandwidth costs, and the impact on the
battery life of the devices.
%
%
\Cref{tab:dongle-ops} shows the compute, bandwidth, and user involvement
characteristics of the three main operations performed by the devices:
(i) {{\ephID} collection}, (ii) encounter history upload, and (ii) risk
calculation for a user.
Given these characteristics,
we care about the latency of (ii), since it impacts the overall
timeliness of risk dissemination to other users, latency and bandwidth of
(iii), since
it requires high computation and bandwidth, and we care about power consumption
of (i) and (ii), since they have the maximum impact on the device's
battery life.
In \S\ref{subsec:download-latency} and \S\ref{subsec:battery-life}, we present
an evaluation of the latency and energy costs
of realistic implementations of beacons and low-end dongles.
In \S\ref{subsec:deployment}, we describe our experience from a pilot deployment
of {\sys} with BLE beacons and user devices implementing a basic, functional
subset of all the features described in our design.

\subsection{Risk estimation latency}
\label{subsec:download-latency}
\begin{figure}[t]
    \centering
    \includegraphics[width=\columnwidth]{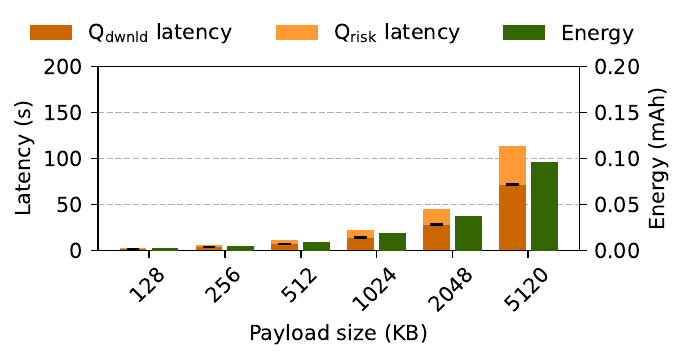}
    \caption{Risk payload sizes vs end-to-end risk dissemination latency and
    energy consumption.~128KB = \update{32768} ephemeral ids, \update{576} 250B
    sized BLE packets.
    }
    \label{fig:download-latency}
\end{figure}
Although {\sys} supports smartphones, our evaluation for personal risk
estimation focuses on dongles, which have fewer compute,
storage, bandwidth, and power resources, and thus present a more
challenging performance target.

We measure the end-to-end latency for risk notification and
dongle power consumption when downloading risk payloads of
various sizes using the broadcast and the active querying protocols.
In all experiments, a dongle and a network device are placed
2m apart in an indoor, barrier-free environment.
The numbers are averaged over three runs and the variance observed is
less than 1\%.

\Cref{fig:download-latency} shows the latency involved in
downloading risk payload (${\rm \bf Q_{dwnld}}$)
and for estimating and notifying the user~of any matches (risk) as a function of
different payload sizes (${\rm \bf Q_{risk}}$).
The total risk estimation latency constitutes \update{$\sim$37.3\%} of the
end-to-end latency.
This latency is dominated by lookups of each of a dongle's
encounter entry in each {\CF} chunk downloaded (see streaming lookup in
\S\ref{sec:prototype}).
The latency of decrypting and XOR-ing the PIR query
results is negligible compared to the {\CF} lookups.
Similarly, uploading a query includes secret-share generation,
encryption, and uploading the PIR queries. The querying latency is a
constant \update{2.3s} regardless of the risk payload size, and is
\update{at most 25\%} of the end-to-end latency.
\update{Finally, the query executes on the {\pirDB} in $\sim$170ms in the
backend, which is negligible compared to the dongle's compute and communication
latency.}

The green bars in \Cref{fig:download-latency} show the energy consumption for
risk notification for different risk payload sizes.

\subsection{Battery lifetime}
\label{subsec:battery-life}

We measured the current consumption of BLE beacons and dongles using Simplicity
Studio's energy profiler, and here we estimate the battery life of the devices.
Again, we do not analyze the battery life of smartphones, since they are already
provisioned with
much higher battery capacity.

\textbf{Beacons.}
The beacons transmit on BLE channels with a power of -10\,dbm or equivalently
0.1\,mW. The average current draw of the beacon is
\update{7\,$\mu$A}, which is dominated by the current draw during BLE
transmission. Thus, with a single coin cell of 220\,mAh capacity, a beacon
can last more than \update{3\,years}.

\textbf{Dongles.}
The dongle's base current draw is \update{0.85\,mA}.
In a \update{60-min} period, the dongle's average current consumption is
\update{0.9\,mA} when only scanning for beacon {\ephIDs}, and \update{1.04\,mA}
when additionally downloading risk information from a
network beacon once using either periodic broadcast or connection-oriented
communication.
Thus, with a coin cell of 220\,mAh capacity, the dongle is expected to last
\update{$\sim$8.8\,days}. Note that this is a
conservative estimate, since users would download risk information only
infrequently, \eg once a day.
\update{With rechargeable cells of even 60\,mAh capacity,
the dongles can last $\sim$2.7\,days on a single charge. This is
practical, since users can be asked to place their dongles on a (wireless)
charger overnight.}

\subsection{Deployment}
\label{subsec:deployment}

{Our prototype does not yet support an end-to-end implementation of the upload
protocol and the PIR protocol for risk querying.
However, our phone app supports uploading a user's encounter history
directly to the backend and downloading the complete risk information from the
backend, thus enabling a pilot deployment.

We tested the end-to-end functionality of {\sys} using a pilot deployment in the
university building over 16 days\footnote{We received university IRB approval
for the deployment.}. We simulated infected users to
trigger uploading of encounter histories and risk dissemination.
The backend server was hosted in a single core Ubuntu 20.04.3 LTS VM with
1GB RAM and 32GB disk. We hosted users' data in a MySQL DB v8.0.27.
We placed 8 BLE beacons, one each in various meeting rooms, labs, and
social areas, and 2 network beacons in a subset of these spaces
for risk broadcast.  We involved 15 volunteers, 10 of whom carried a dongle and
the rest used our smartphone app.
The app users were asked to upload their encounter history to the backend
on three random days. All users checked their simulated exposure risk
everyday. Their devices downloaded the risk information over periodic broadcast
channel only and recorded the number of encounters matched with the
broadcast. App users saw the number of matched entries on their phone
screens and the dongles' LEDs blinked faster to indicate non-zero 
matches.

\textbf{Statistics.}
Over the period of the experiment, the beacons generated a total
of 12288 unique {\ephIDs}, and the user devices captured a total of
11670 of these {\ephIDs}. When the app users uploaded their
encounter history, they uploaded an average of 155 {\ephIDs} to the
backend. The number of other participants whose devices found matching entries
for each of the three uploads was 8, 6, and 5.

\textbf{User experience.}
Our users found both the dongles and the app intuitive and easy to use. In
the future, the smartphone app could provide better visualization of the
encounter data. The dongles~could also be allowed to pair with a personal
device, which the user is willing to trust, to provide similar visualizations of
data as the phone~app.

\paragraph{Evaluation summary.}
Our results indicate that {\sys} can be practically deployed with low
infrastructure and maintenance costs, and can be easily adopted by users. {Both
smartphones and
low-end dongles can upload encounter data and receive risk information of their
regions of interest with modest bandwidth, latency, and energy costs.}

%% file: discussion.tex
\section{Discussion}
\label{sec:discussion}

\textbf{Beacon placement.}
The density and placement of beacons is important for minimizing false negatives
and false positives in {\sys}.
False positives arise when a user receives a risk notification even though they
have not been in close contact with an infected user. For instance, a false
notification may be generated when two users encounter a beacon placed on a
glass door but from opposite sides of the door.
False positives~can~be reduced by placing a sufficient number~of
beacons in a location and using well-known localization
techniques~\cite{naghdi2019trilateration}, and by
relying on the beacons' descriptors encoding information, such as
temperature, humidity, etc.

False negatives arise when potential transmissions between users are missed, for
instance, because of users meeting in locations where there are no beacons.
Beacon deployments can be planned strategically to minimize false negatives. For
instance, restaurants are likely to be more crowded than parks; therefore,
restaurants must be prioritized in a partial rollout.

\textbf{False positives in non-contemporaneous events.}
If Alice and Bob (who is infected) visit a beacon in the same epoch and Alice
leaves before Bob's visit, she would still receive a risk notification for this
beacon visit, even though she was not exposed to Bob.
To eliminate such false positives, {\sys} could additionally
transmit in the risk information the beacon's~start timestamp and interval as
observed in the infected user's encounter with the beacon. For the matched
{\ephIDs}, \update{user devices} would then compare the beacon's timestamp and
interval recorded in their own log and in the risk information for overlap. A
user would be
notified only if the time interval in \update{device}'s log overlaps with and
starts later than the interval in the risk information.

\textbf{Interoperability with existing {\ct} systems.}
{\sys}'s beacons can broadcast ephemeral IDs compatible with the
Google/Apple Exposure Notification (GAEN) protocol used by most
{\spects}~\cite{gaen-api}, allowing the beacons to seamlessly interoperate with
deployed apps.

{\sys} can also achieve bidirectional interoperability with manual
contact tracing~\cite{pancast-simulations-anon}.
Health authorities may manually obtain location data from consenting diagnosed
individuals and insert records into {\sys}.
Thus, even users who do not carry a dongle can contribute to subsequent risk
estimates and broadcasts.
Conversely, by providing a user-comprehensible record of visited beacon
locations, {\sys} can be used as a diary aiding~the memory of individuals
participating in manual tracing. Finally, since {\sys} can associate
risk events with locations, information about potential superspreading events
can be broadcast using traditional means of communication.

%% file: related.tex
\section{Related work}
\label{sec:related}

We discuss digital {\ct} systems that use various technologies, such
as Bluetooth, GPS, or QR codes, to track users' trajectory and/or proximity to
other users. We also discuss privacy-preserving techniques relevant for risk
information retrieval.


\textbf{P2P {\ct} systems.}
{\spects} record
instances of physical proximity with devices of other individuals via
close-range Bluetooth exchanges between user devices.
%
%
Centralized {\spects}~\cite{pepp-pt-ntk, bay2020bluetrace, opentrace,
tracetogether} collect and manage user data centrally, placing a
high degree of trust in the central authority. Decentralized
{\spects}~\cite{dp3t-whitepaper, chan2020pact, tcn, contra} minimize data
collection 
to preserve privacy, but this
prevents aggregation of data for~epidemiology.
{\sys} facilitates analysis by enabling collection of contextual
information with encounters.

In {\spects}, users' devices actively transmit messages and, thus, are
vulnerable to eavesdropping and~surveillance
attacks~\cite{vaudenay2020centralized}. Furthermore, an attacker could relay or
replay captured ephemeral ids~\cite{vaudenay2020centralized,
baumgartner2020mind}. {\sys} overcomes these attacks because users' devices
mostly listen passively, and beacons' location-time configurations can be
corroborated with external trusted sources.


\textbf{P2I {\ct} systems.}
Reichert et al.~\cite{reichert2021lighthouses} propose an architecture where all
beacons (``lighthouses'') and user devices are smartphones.
Lighthouses collaborate with the backend for removing false positives in risk
notification to a user who left a beacon before an infected user visited it (see
\S\ref{sec:discussion}). Consequently, users need to~trust the lighthouses to
not collude with the backend in leaking their data. {\sys}
eliminates the false positives in risk notification without requiring
collaboration between the backend and the network beacons, which may be operated
by untrusted third parties. Unlike~lighthouses, {\sys} can also handle relay/replay
attacks.

{\pancast}~\cite{pancast-simulations} uses Bluetooth beacons and supports
dongles similar to {\sys}. However,
{\pancast} focuses on evaluating---through simulations---the
effectiveness of a beacon-and-dongle architecture for risk notification, and the
benefits of interoperating with manual tracing. {\sys}, on the other hand,
builds a real system using \update{smartphones and} low-end IoT devices,
addressing technical challenges
in achieving security, scalability and performance efficiency.
{\pancast} assumes a pure broadcast-based risk dissemination architecture,
which can be very expensive in terms of bandwidth and even latency during
high infection rates.
{\sys} uses an IT-PIR based active querying protocol, thus minimizing bandwidth,
latency, and power costs for \update{user devices}.

Systems that use QR codes
\cite{crowdnotifier,safeentry-qr-singapore,canatrace-qr-canada,hoffman2020towards}
rely on static QR codes for each registered location.
Static codes allow linking a user's multiple visits to the same locations, thus
revealing more information about their location history.
%
Other applications track location history using
GPS~\cite{mit-safepaths,pathcheck-gps-us}, which is imprecise and invasive, or
using encounters with WiFi access points~\cite{trivedi2020wifitrace}, which
requires infrastructure that is relatively expensive compared to {\sys}'s
infrastructure.

\textbf{Privacy-preserving risk querying.}
%
{\sys}'s PIR technique with dynamic block sizes is similar to
LBSPIR~\cite{olumofin2010lbspir}.
However, LBSPIR was designed for smartphone applications, which can retrieve the
dynamic block layout of the PIR DB from the server and then adapt the size of
each PIR query based on users' privacy preferences.
{\sys} relies on
a hierarchical geographical tiling, which enables PIR with minimal interaction
between the server and a dongle, and provides a fixed privacy guarantee with
fixed overheads for all queries.

EpiOne~\cite{trieu2020epione} proposes a two-party private-set intersection
cardinality (PSI-CA)
technique, to enable users to find {\em how many} entries in their
encounter history match those of patients.
{\sys} reveals {\em which} location-time entries in a user's history match those in the risk information.
Hence, {\sys} provides more context for a user's exposure risk
without compromising patients' privacy.
Secondly, EpiOne relies on computational PIR, Merkle tree and
zero-knowledge proofs to provide privacy for queriers and the infected
individuals. These mechanisms have high computational and communication costs.
{\sys} provides similar guarantees but using IT-PIR, differential privacy, and
cuckoo filters, which offload most computational costs to the server and thus
{are more suitable for user devices, particularly low-end~dongles.}

%% file: conclusion.tex
\section{Conclusion}
\label{sec:conclusion}

We focus on building a systematic, inclusive, and scalable contact tracing and
risk notification system in preparation for future needs.
%
To this end, we present {\sys}, a novel system for epidemic risk mitigation
based on person-to-infrastructure encounters, showing that it is possible to
extract significant utility without compromising on security.
We presented the design and a prototype of {\sys} along with a detailed analysis
of its security, efficiency, and scalability. We demonstrated {\sys}'s
practicality through a pilot deployment in a university building.
\update{We plan to~evaluate {\sys} in a real-world deployment in
the future.}

